# Stress relaxation microscopy (STREM): Imaging mechanical force decay in cells

Susana Moreno-Flores[1], Rafael Benitez[2], Maria dM Vivanco[3], and José Luis Toca-Herrera[1]

[1] Biosurfaces unit, CIC BiomaGUNE, Paseo Miramón 182, 20009 San Sebastián-Donostia, Spain
[2] Dept. Mathematics, Centro Universitario de Plasencia, Universidad de Extremadura, Avda. Virgen del Puerto 2, 10600, Plasencia, Spain
[3] Cell Biology & Stem Cells Unit, CIC BioGUNE, Parque tecnológico de Bizkaia, Ed. 801A, 48160 Derio, Spain

* Corresponding author:

José L. Toca-Herrera
CIC BiomaGUNE
Paseo Miramón, 182
20009 San Sebastián (Spain)
Tel: ++34 943 00 53 13
Fax: ++34 943 00 53 01
E-mail: jltocaherrera@cicbiomagune.es

**ABSTRACT**
We have developed a novel scanning probe-based methodology to study cell biomechanics. The time dependence of the force exerted by the cell surface on a scanning probe at constant local deformation has been used to extract local relaxational responses. The generalized Maxwell viscoelastic model that accounts for multi relaxations fully describes the mechanical behaviour of the cell surface that exhibits a bimodal relaxation. Within the range of tested forces (0.1-4 nN) a slow and a fast relaxation with characteristic times of 0.1 and 1s have been detected and assigned to rearrangements in the cell membrane and cytoskeleton cortex, respectively. Relaxation time mapping allows to simultaneously detect non-uniformities in membrane and cytoskeletal mechanical behaviour and can be used as both identifying and diagnosing tools for cell type and cell disease.

INTRODUCTION
Cell biomechanics is becoming a diagnostic tool in cell biology. Cell response to mechanical stimuli are distinctive of cell type,[1] cell function[2] and cell damage,[3] and are believed to be strongly dependent on the cytoskeleton.[1] Structural proteins such as keratins especially confer mechanical resistance on epithelial cells;[4] human diseases like most of the haemolytic anemias and cirrhosis are associated with elasticity loss in red blood cells and hepatocytes, respectively. Additionally, carcinoma cells exhibit anomalous compressibility or elasticity when compared to healthy cells of the same type.[5,6,7]

Few are the ways to impose a mechanical stimulus to a cell and observe its response:

In the so-called transient experiments sudden stress induce creep in cells that can be monitored using optical microscopy. Based on that, techniques such as micropipette aspiration[8] and microplate manipulation[9] are particularly relevant in the study of whole-cell creep mechanics of non-adherent, light-adherent or suspended cells. Additionally, dynamic shear stress induces time-dependent deformations that can be detected in microrheology (i.e. dynamic) experiments. In this case, laser-track[10] and magnetic probe-based[11,12] techniques allow to study the viscoelasticity of the intracellular space and the cell surface by monitoring the displacement of internalized and surface-attached microparticles, respectively.

Of all the different techniques currently available to test cell response to mechanical stimuli, the cell poker[13] and scanning probe (SP) based-force spectroscopy[14] are most suitable to study local behaviour on adherent cell surfaces and tissues.[2] Deformation and stress are applied normal to the cell surface either with a microsized glass stylus or a submicro-sized silicon(nitride) tip positioned at a certain location. Due to the smaller tip size, the SP-based technique provides better spatial resolution and has been mainly applied to the study of the elastic stress-strain behaviour of cells and the obtention of elastic moduli.[15] However, modelling has been greatly restricted in these cases. Except for SP- based microrheological studies,[16] cells have been conceived as purely elastic, homogeneous materials that are subjected to small (10-100 nm), sudden deformations. Typical probing areas are submicrometer-sized, which questions the validity of cells being homogeneous bodies normal to and along the cell surface. The scanning probe can sense various cell components, especially at high deformations. All these components (cell membrane, actin cortex, other cytoskeletal components) may respond differently to probe-induced stimuli. Additionally, probing the cell at different locations provides information on the lateral distribution of mechanical responses, which may in turn differ.

In this work we describe an SP-based imaging methodology that generates stress relaxation maps of complete cells, which have been subjected to larger range of deformations than those reported to date with this technique. Our data analysis takes into account both cell three-dimensional heterogeneity and cell viscoelasticity. To test our method, we have used human breast cancer cells (MCF-7). This epithelial breast adenocarcinoma cell line is one of the most frequently used model systems to study breast cancer. The MCF-7 cell line was derived from a pleural effusion of a patient with metastatic breast cancer,[17] however, this line normally does not metastasize. The relaxation of the force exerted by MCF-7 cells on SP-cantilevers after the application of a sudden deformation has been characterized within a wide range of applied deformations and cell locations. A multicomponent viscoelastic model has been used to interpret the data, to extract mechanical properties and to correlate the latter to cell topology.

**EXPERIMENTAL METHODS**

*Sample preparation*. MCF-7 cells were grown at 37ºC and 5% $CO_2$ in Dulbecco's modified eagle medium (DMEM, Sigma) supplemented with 8% fetal bovine serum (FBS, Sigma), 2% 200mM L-glutamine, 0.4% penicilline/streptomicine (PEN/STREP, Sigma). For force measurements, the cells were subcultured on borosilicate glass coverslips (diameter 24 mm and 0.16 mm thickness) at a density of 25K/ml, 15K/ml and 10K/ml and left to incubate for 1,2 and 3 days, respectively. Prior to force measurements, the cells were washed in $CO_2$-independent cell medium (Leibowitz medium, L15, Sigma) and measured in the same medium at 37ºC.

*Force-time curves.* Measurements were carried out on different cell clusters for the same sample with a Nanowizard II (JPK Instruments, Germany) coupled with a transmission optical microscope (Axio Observer D1 Zeiss, Germany). Uncoated SiN cantilevers of nominal spring constant of 0.06 N/m (MLCT, Veeco Instr., USA) were used for both cell imaging and force measurements. The cantilevers were previously cleaned in acetone and ethanol to remove impurities and their spring constants evaluated by the thermal method. The cell-coated glass substrates were then mounted in a low-volume cell incubator (Biocell, JPK Instruments, Germany), with 400 µl L15 cell medium and thermalized at 37ºC. Individual force-time curves were recorded at a speed of 5µm/s and at maximum loads of 0.5,1,2,3 and 4 nN on different cells and at different cell positions. Force relaxation was registered at constant height mode, where the contact time was set to 2 seconds. Together with the force relaxation, approach and withdrawal curves were also obtained. Cell deformations were obtained from the corresponding approach curves by computing the vertical displacement of the probe between the contact point and the maximum applied load, in other words, the extension of the contact, non-zero force region. This amount was substracted from the cantilever deflection, which was likewise computed from approach curves performed on the glass substrate under the same applied load. For STREM, the evaluated area was divided either in 25x25 or 30x30 pixels and in each pixel a force-time curve was registered. The maximum load for mapping was kept constant to 2 nN. In all cases, optical micrographs of the cells were taken before and after the experiment to rule out possible tip-induced cell damage. A self-developed analysis software was used to extract the normal tension decay, the cell deformation and the relaxation time during contact time for each pixel and construct the 2D maps.

**RESULTS**
*Force-time curves – compressive force relaxation curves*
We have registered as a function of time the force exerted by MCF-7 cells (figure 1a) on the SP cantilever during tip approach, tip-cell contact, and tip withdrawal (figure 1b). At $t_0$ starts the contact time, which is the time when the tip remains in contact with the cell, and the force steadily decreases with time from an initial value (the maximum load, see figure 1b). The force eventually reaches a plateau if contact is maintained sufficiently long (curve 1 in figure 1b). This observed force decay is symptomatic of probing non-elastic bodies, in opposition to purely elastic ones like the glass substrate, which does not exhibit such behaviour (curve 2 in figure 1b). We have additionally observed that the force-time dependence in MCF-7 cells - under the studied loads - obeys a double-exponential decay (dashed line in figure 1b).

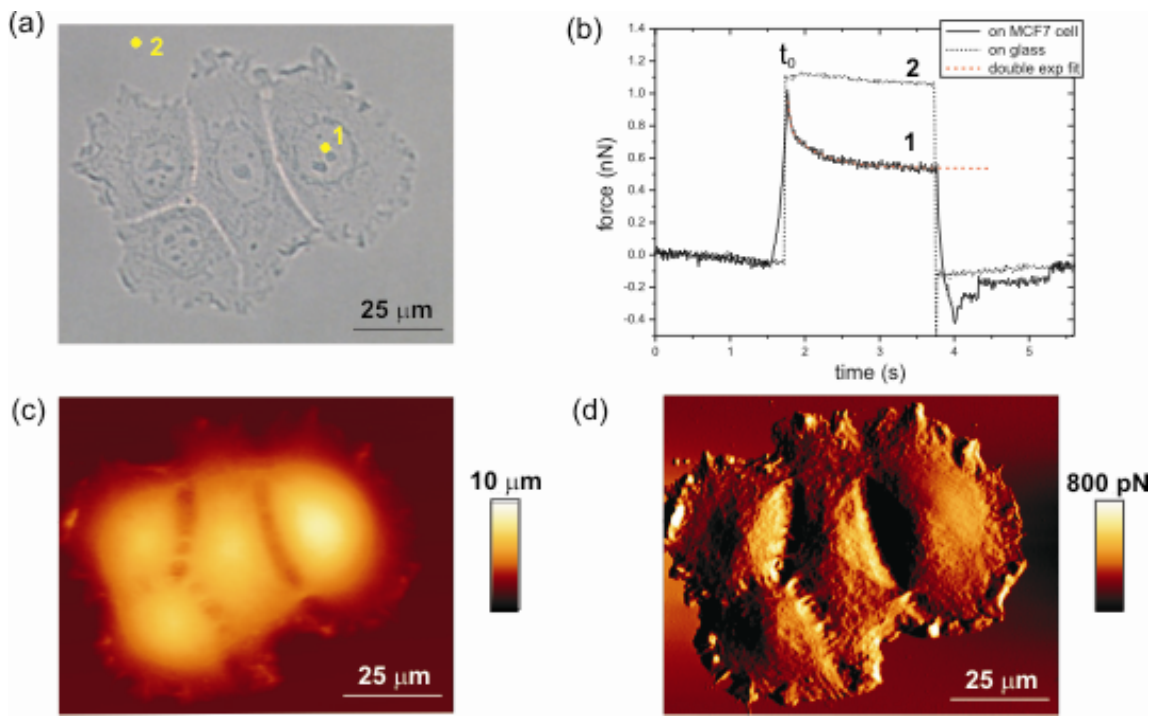

**Figure 1.** a) Differential interference contrast (DIC) optical micrograph of a cluster of four MCF-7 cells where nuclear regions are distinguishable; b) force-time curve registered during tip approach (times smaller than $t_0$), tip-cell contact ($t_0$ - 4s) and tip withdrawal (times longer than 4s). Curves are performed on the nuclear region of one of the MCF-7 cells (location depicted as 1 in figure 1a, curve 1) and on the glass substrate (location depicted as 2 in figure 1a, curve 2); c) height SP micrograph of the same cluster indicates that the maximum cell heights are attained on the nuclear regions ($\geq$ 5 $\mu$m); d) SP cantilever deflection micrograph showing the details of cytoskeletal fibers at cell edges.

An exponential force decay is the typical relaxation response of certain linear, isotropic viscoelastic bodies subjected to a constant deformation.[18] According to the spring-dashpot model developed by Maxwell, which consists of an elastic spring connected in series with a viscous dashpot, a material characterized by a compressive elastic modulus E, and a viscosity $\eta$ exhibits a force response to a sudden and constant deformation set at time $t_0$ that can be defined as follows:

$$F = A\exp\left(\frac{-(t-t_0)}{\tau}\right)$$
$$\tau = \frac{\eta}{E}$$
(1)

where A is the force amplitude of the relaxation and $\tau$ the relaxation time. This relation holds as long as both the deformation and the contact area (region of the material along which the mechanical deformation is applied) are constant.[19] Accordingly, a generalized Maxwell model consisting on N parallely arranged Maxwell elements describes multiexponential decays in heterogeneous materials as follows:

$$F(t) = A_0 + \sum_{i=1}^{N} A_i \exp\left(\frac{-(t-t_0)}{\tau_i}\right)$$
$$\tau_i = \frac{\eta_i}{E_i}$$
(2)

where $A_0$ accounts for the instantaneous (purely elastic) response. Each element thus contributes to the overall response as an individual relaxational process. In our experiments the cantilever position was kept constant during the contact time and the forces and deformations are applied normal to the cell surface. Contact area is assumed to be constant along with deformation. Therefore the force change we register on cells during this time is thus interpreted as force relaxation under constant compression with N=2 simultaneously-occurring processes. $A_1$ and $A_2$ are the corresponding decay amplitudes and $\tau_1$ and $\tau_2$ their relaxation times. On the glass substrate, no noticeable force relaxation is detected and thus is characterized by zero decay amplitudes ($A_1=A_2=0$).

*Dependence of the relaxation on maximum loads and subcellular localisation*
When performing the experiments at different loads from 0.5 to 4nN, the time dependence of the force during the contact time always decays in a double-exponential fashion. Figure 2a shows as black traces the experimental curves performed on the nuclear region of an MCF-7 cell and the corresponding double-exponential fitting (red curves) according to equation 2. The quality of the fits is good as can be seen in figure 2a, which allowed to obtain both the overall amplitude of the force decay ($A_1+A_2$) and the relaxation times ($\tau_1$ and $\tau_2$) in each case. Analogous experiments on the cytoplasmic region of the cells allowed comparison of relaxational processes set at different subcellular localisations. The result of this comparison is shown in figures 2b and 2c, where the amplitude of the force decay and the relaxation times are plotted against the cell deformation (figure 2b) and the initial load (figure 2c), respectively.

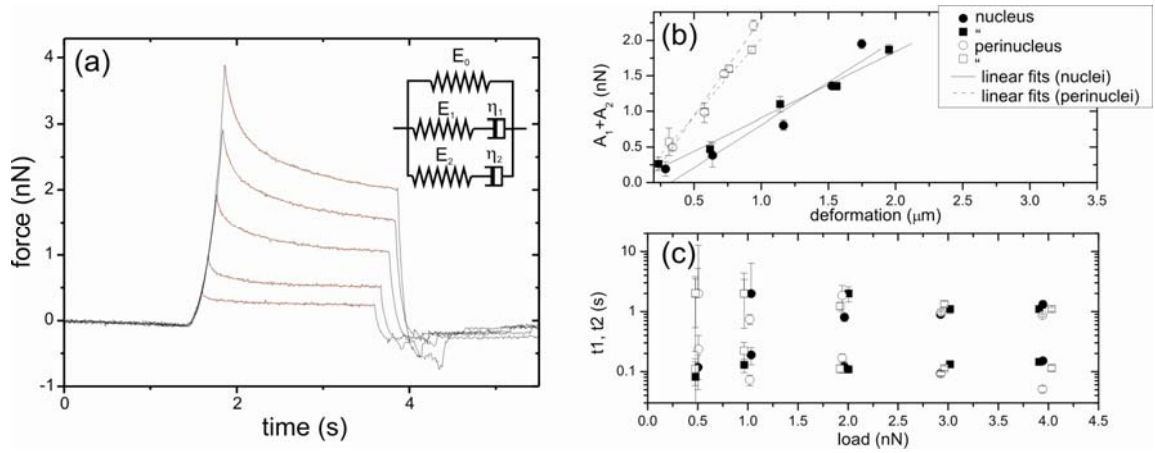

**Figure 2.** a) Force-time curves (black traces) on position 1 (see figure 1) at different maximum loads (0.5, 1, 2, 3 and 4 nN), the red curves being the double-exponential fittings. Inset: viscoelastic model that consists of a spring connected in parallel with two Maxwell elements ; b) total force decay versus local deformation for two nuclear and two cytoplasmic (perinuclear) regions of the same cell, the straight lines are fits to the experimental data; c) relaxation times, $\tau_1$ and $\tau_2$, for those regions.

The behaviour of the force decay-to-cell deformation differs on the nuclear and cytoplasmic (i.e., perinuclear) regions of the cell. Figure 2b shows a linear dependence on both regions within the range of maximum loads evaluated; however, the nuclear region is prone to larger deformations than the perinuclear region, and therefore the data of the former extends to higher deformation values. Hence, force decay-to-cell deformation ratio (i.e., the normal tension decay) is larger on the perinuclear region than on the nuclear region, which denotes that this magnitude depends on the subcellular localisation. Figure 2c, shows that the fast and slow relaxation times, $\tau_1$ and $\tau_2$ respectively, differ one order of magnitude, $\tau_1$ being in the range of 100 ms and $\tau_2$ in the range of seconds. Surprisingly their values neither depend substantially on the maximum load applied nor on the subcellular localisation. However, evaluating at random positions does not provide a complete picture of the cell response as we will show in the next section.

*Mapping compressive force decays and relaxation times: STREM images*
Magnitudes that can be extracted on each cellular location, such as normal tension decays, $\tau_1$ and $\tau_2$, can be mapped to obtain a three dimensional view of the relaxational processes on the complete cell surface. In this case, the maximum load applied is unique for all positions, and force-time curves are taken while the SP cantilever is step-scanning an individual cell or a cell cluster. As an example, figure 3 shows STREM maps of two individual MCF-7 cells. The maximum load has been set equal to 2 nN, high enough to clearly distinguish both relaxations and low enough to be within the linear regime (see figure 2b) and ensure cell integrity. Distribution of cell heights accounts for image contrast in figure 3a, where nuclear regions appear thick and bulgy while perinuclear regions appear flatter.

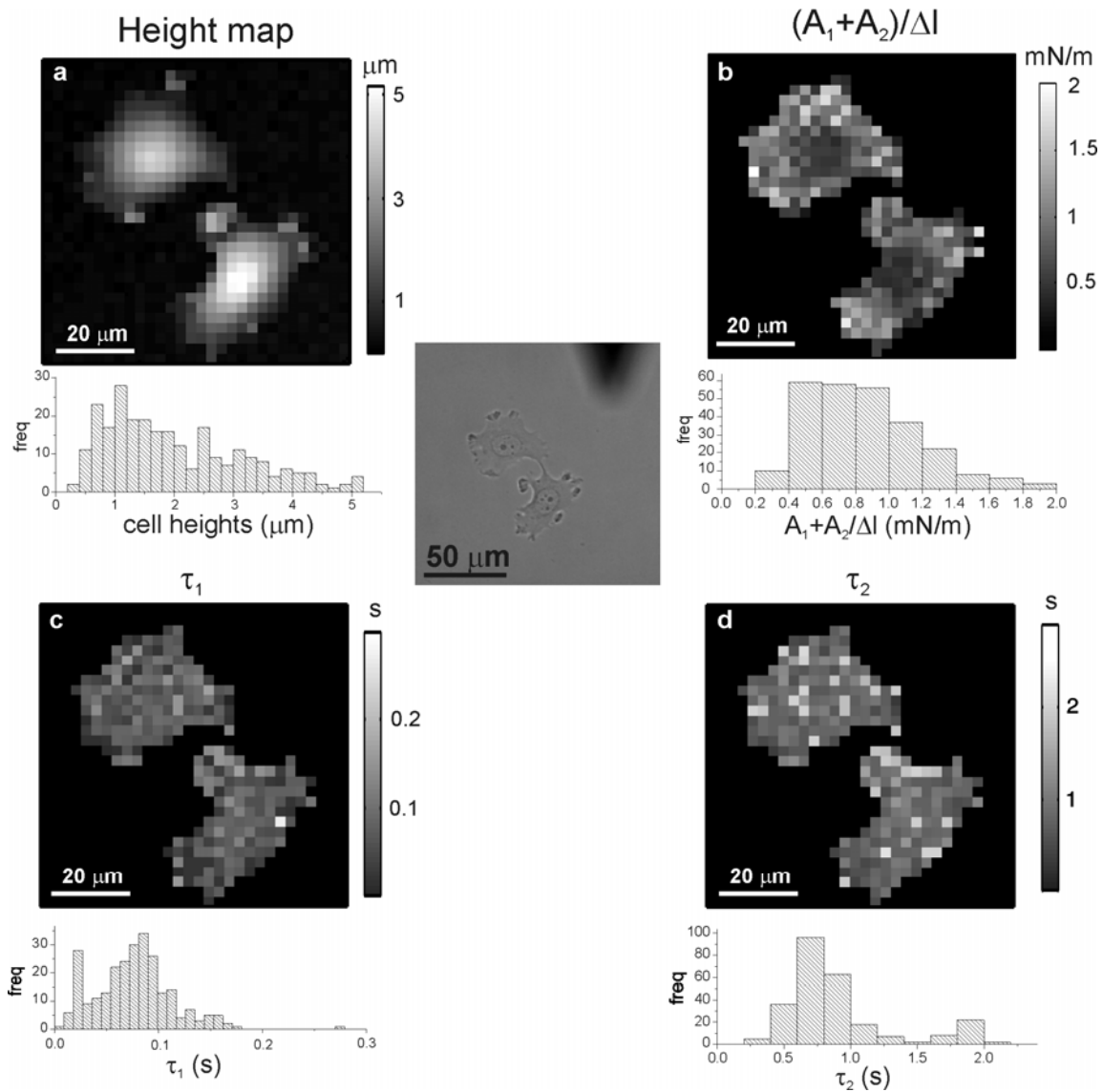

**Figure 3.** Height (a) and STREM (b-d) images of individual MCF7 cells with the corresponding histograms. The height map is a distribution of heights at constant force (1 nN), cell heights extend from 1 um (perinuclear regions) to 5 um (nuclear regions); b) map and histogram of the ratio between force decay and deformation, the normal tension decay; c) map and histogram of the fast relaxation time; d) map and histogram of the slow relaxation time. Histogram binning (i.e. width of columns) is comparable to the parameter error. Black pixels outside the cells refer to substrate, where no relaxation or deformation occurs (no numbers associated).

Height SP micrographs (figure 2c) on MCF-7 clusters confirm this fact. The force decay amplitudes in figure 3b have been divided by the cell deformation at each pixel to obtain normal tension decays and to allow comparison between the different cell localisations. Image contrast in figure 3b is thus indicative of biomechanical disparity and it correlates with cell morphology: the bulgy, nuclear regions of the cells exhibit larger deformations and thus appear darker than the perinucleus. Figures 3c and 3d show maps of the fast ($\tau_1$) and slow ($\tau_2$) relaxation times, respectively. These measurements illustrate the importance of mapping relaxation times of the whole cell in opposition to the results shown in figure 2c, where measurements were performed on a few cell locations. Figure 2c would lead to the erroneous conclusion that the cell time response is positionally homogeneous. On the contrary figures 3c and 3d show the complexity of the cell response, which is positionally non-homogeneous and more comprehensively characterized by mapping and histograms.

## DISCUSSION

### Overall force decay vs deformation

Overall force decay depends linearly on cell deformation in both nuclear and perinuclear regions (figure 2b). However, the perinuclear region is usually less compliant to deformation than the nuclear region,[14,20] which is also observed in our experiments. Therefore the normal tension decay $(A_1+A_2)/\Delta l$,[21] the slope of the overall force decay versus deformation, depends on subcellular localisation. On perinuclear regions, the slopes are approximately twice as much as those obtained on nuclear regions (figure 2b).[22] This numerical discrepancy may be strongly linked to the relative higher compressibility of the nuclear region in comparison to the perinuclear. The former has a higher content of intracellular fluid and the volume density of cytoskeletal fibers is lower than on the perinucleus, which accounts for its higher compressibility. Mapping normal tension decays on a complete cell is thus most illustrative of cell morphology and response distributions (figure 3b).

### Bi-exponential decay: possible reasons

The observed biexponential decay may have two possible reasons. Upon cell compression, compressive and shear forces may occur, which result in two relaxational processes if decoupled. Alternatively, it is reasonable to expect that two relaxations may also account for probing two different cell environments.

Let us first consider the first case. Our probe has a pyramidal shape. On contacting the cell, the local cell shape may be deformed around the pyramidal tip. In this case, indentation would take place and both compressive and shear deformations occur upon the contact area that extends beyond the apex of the tip (compressive) to the pyramidal sides (shear). Hence both compressive and shear forces would contribute to the detected relaxation. The two relaxational processes found may thus account for the existence of these two types of forces:

$$F(t) - A_0 = F_{compressive}(t) + F_{shear}(t) = A_{compressive} \exp\left(\frac{-(t-t_0)}{\tau_{compressive}}\right) + A_{shear} \exp\left(\frac{-(t-t_0)}{\tau_{shear}}\right) \quad (3)$$

each one characterized by a decay amplitude and a relaxation time. In this case, the contribution (i.e., amplitudes) of these two terms to the overall force decay should certainly depend on probe geometry. A pyramidal-nanosized tip indents more than a microsized spherical colloid and therefore should induce a higher shear force.

The contribution of the second term in equation 3 to the overall force decay, $A^*_{shear} = A_{shear}/(A_{tcompressive}+A_{shear})$, should thus be greater in the case of a pyramidal probe than in the case of a colloidal probe. In other words, $A^*_{shear}$ should be larger for tips than for colloids and $(A^*_{shear})^{(tip)} / (A^*_{shear})^{(colloid)}$ larger than one, irrespectively of the applied load. In our experiments we obtain two amplitudes, $A^*_1 = A_1/(A_1+A_2)$ and $A^*_2 = A_2/(A_1+A_2)$, which cannot *a priori* be attributed to shear or compressive force decays. We thus performed stress relaxation experiments with pyramidal and colloidal tips on nuclei of MCF-7 cells and plotted $(A^*_1)^{(tip)}/(A^*_1)^{(colloid)}$ and $(A^*_2)^{(tip)}/(A^*_2)^{(colloid)}$ as a function of the applied load (figure 4).

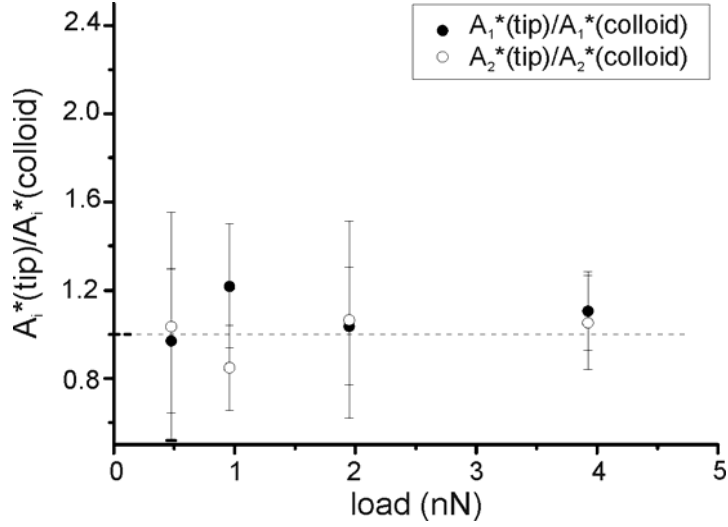

**Figure 4**. Force decay amplitude ratios for different probe geometries. Tip: silicon nitride square pyramid with 100 nm apex radius. Colloid: silicon bead of 8 μm diameter. The magnitude $A_i^*$ (i=1,2) refers to the ratio $A_i/(A_1+A_2)$. Within the experimental error both amplitude ratios are indistinguishable from one. The experiments were performed on the nuclear area of MCF-7cells. Results show statistical averages from 13 (pyramidal tip) and 10 (colloidal probe) cells.

The results show that, within the experimental error, the calculated ratios are close to one, meaning no difference in the contributions to the overall force decay. The biexponential decay is therefore not due to the existence of decoupled shear and compressive relaxational processes.

In the second case, when shear relaxation does not occur or cannot be decoupled from compressive relaxation, the two observed decays may account for relaxational processes in two different cell environments and thus the equation has the following form:

$$F(t) - A_0 = F_{cc1}(t) + F_{cc2}(t) = A_{cc1} \exp\left(\frac{-(t-t_0)}{\tau_{cc1}}\right) + A_{cc2} \exp\left(\frac{-(t-t_0)}{\tau_{cc2}}\right) \quad (4)$$

where *cc1* and *cc2* stand for cell component 1 and 2 respectively. The time scale of the fast relaxation process in our experiments, $\tau_1$, lies between 0.1 and 0.3s, which greatly resembles the response times of red cell membranes[21] and of macrophages (t=0.218s).[23] The slow relaxation, $\tau_2$, is of the order of seconds, which is within the time scale of cytoskeletal rearrangements.[10,24] Assigning the fast and slow relaxations to membrane and cytoskeletal responses may be *ad hoc*; however it agrees with the fact that both relaxations proceed with similar time scales on both nuclear and perinuclear regions of the cell (figure 2c). The two outermost components that completely surround the cell are the cell membrane and the cytoskeletal cortex, [25,26] which are most likely sensed by the scanning probe.

STREM images show that it is possible to exploit the capabilities of the scanning probe technique to address local mechanical processes within the cell and obtain relaxation maps. The values of $\tau_1$ and $\tau_2$ are similar along their surface, though not identical. Non-uniformities in the relaxation time maps appear randomly distributed in the case of the $\tau_1$ map and temptatively addressed to local structural "anomalies" on the cell surface (e.g., protein clusters, gap junctions, local adhesion points[27]). In the $\tau_2$ map a double distribution of relaxation times was easily observed, the higher values are mainly found in the regions close to the cell edges, where the volume density of fibers are high (see figure 2d). STREM can thus provide access to regions within the cell characterized by distinct time responses.

**CONCLUSIONS**

We have developed an SP-based methodology accessible to an alternative set of local mechanical parameters that through mapping, characterize complete cells. We have used the MCF-7 cell line as a model system to which we have applied local deformations with an SP cantilever. The force decay has been fully characterized as a function of time. Cells relax according to two simultaneously occurring processes that may involve cell membrane and cytoskeletal rearrangements. Our work illustrates the importance of STREM mapping in the biomechanical characterization of cells. Thorough mapping of normal tension decays and time relaxations of complete cells provides a comprehensive view of the complexity of cell relaxational biomechanics and eases to link the local mechanical behaviour with cell morphology. As a matter of fact, STREM is a potential tool to localise gap-junctions and caveolae in *invitro* cells.

The authors envisage STREM as a developing technique with wide applicability though it requires refinement. As a matter of fact, the influence of the elasticity of the scanning probe on the observed relaxations is still an open question. Diagnosing cell activity and dysfunction are among the biomedical applications of STREM, although its utility mightnot restricted to cells. (Bio)polymer films and scaffolds, liposomes and tissue among others are also viscoelastic materials that can be fully characterized by this method.

**ACKNOWLEDGMENTS**

The authors thank Verónica Saravia for technical assistance in the experiments with colloidal probes and also to Marco Piva for help with cell culture. We additionally thank Dr. Francisco J. del Castillo, Dr. Kathrin Melzak and Prof. Helmuth Moehwald for reading and useful comments. SMF, MdMV and JLTH thank the ETORTEK programme of the Basque Government for financial support.